# Observation of half-integer thermal Hall conductance

Mitali Banerjee[+], Moty Heiblum[+,*], Vladimir Umansky[+], Dima E. Feldman[++], Yuval Oreg[+], and Ady Stern[+]

[+]Braun Center of Sub-Micron Physics, Department of Condensed Matter Physics, Weizmann Institute of Science, Rehovot, Israel 76100

[++]Department of Physics, Brown University, Providence, RI 02912, USA

([*]Corresponding author: moty.heiblum@weizmann.ac.il)

**Topological states of matter are characterized by topological invariants, which are physical quantities whose values are quantized and do not depend on details of the measured system. Of these, the easiest to probe in experiments is the electrical Hall conductance, which is expressed in units of $e^2/h$ ($e$ the electron charge, $h$ the Planck's constant). In the fractional quantum Hall effect (FQHE), fractional quantized values of the electrical Hall conductance attest to topologically ordered states, which are states that carry quasiparticles with fractional charge and anyonic statistics. Another topological invariant, which is much harder to measure, is the thermal Hall conductance, expressed in units of $\kappa_0 T = (\pi^2 k_B^2/3h)T$ ($k_B$ the Boltzmann constant, $T$ the temperature). For the quantized thermal Hall conductance, a fractional value attests that the probed state of matter is non-abelian. Quasiparticles in non-abelian states lead to a ground state degeneracy and perform topological unitary transformations among ground states when braided. As such, they may be useful for topological quantum computation. In this paper, we report our measurements of the thermal Hall conductance for several quantum Hall states in the first excited Landau level. Remarkably, we find the thermal Hall conductance of the $v$=5/2 state to be fractional, and to equal $2.5\kappa_0 T$.**

The even-denominator fractional quantum Hall state in the first excited Landau level at $v$=5/2 has been a subject of intense research for the past thirty years [1]. After its first observation [2], it was bravely suggested that the state may be a manifestation of p-wave superconducting-like condensation of composite fermions in an effective zero magnetic field [3, 4, 5]. Furthermore, it has been predicted that the state carries quasiparticles whose mutual statistics is non-abelian [4]. Consequently, the ground state of several quasiparticles remains degenerate even at their fixed positions; making them attractive for topological quantum computing [1]. Yet, this prediction has been hard to test experimentally, since the relatively easy accessible experimental probes, such as electric response functions, do not reflect the topological order of the state. Even with the demonstration that the state's quasiparticle charge is $e^*=e/4$ [6, 7], and the observation of a topologically protected *upstream* propagating neutral mode [8], a family of possible abelian as well as non-abelian topological orders are still viable candidates for the $v=5/2$ state (see Table in the Extended Data).

The thermal Hall conductance is a unique probe that can distinguish clearly between the different candidate topological orders. Non-abelian states at $v=5/2$ should support half-integer quantized thermal Hall conductance [5]. Here we report an observation of such half-integer quantization.

It is worth giving a brief summary of the different orders predicted for this state. Numerical work lent some support [9, 10] to the non-abelian Pfaffian [4] and anti-Pfaffian topological orders [11, 12]. Among other possible states are the $SU(2)_2$, K=8, 331, and the 113 liquids, as well as their particle-hole conjugates [13, 14, 15, 16]. A wire construction of these states introduces possible generalizations [17]. The non-abelian PH-Pfaffian order was briefly introduced in [12], and was beautifully interpreted in terms of Dirac composite fermions with the particle-hole (PH) symmetry in [18]; with a proposed electronic wave function in [19]. Yet, microscopic PH-symmetry is absent in realistic $v=5/2$ systems, and numerical calculations gave no evidence of a PH-symmetric order. At the same time, it was argued that breaking the particle-hole symmetry may, counter-intuitively, stabilize the PH-Pfaffian order [19, 20].

Moreover, the existing transport data, such as the existence of topological protected upstream neutral mode [8, 21] and the observed non-linearity in tunneling characteristics [22, 23, 24, 25], are consistent [19] with only two of the proposed topological orders: the abelian 113 liquid and the non-abelian PH-Pfaffian states (see Table in Extended Data).

The thermal Hall conductance is defined in a two terminal measurement as $g_Q=dJ_Q/dT=KT$ with $J_Q$ being the heat current, $T$ the temperature, and $K$ the thermal conductance coefficient. This is a highly important property of the system. In particular, for a 1D ballistic channel the maximal value of $K$ is $\kappa_0$, with $\kappa_0=\pi^2 k_B^2/3h$ being a universal constant combining only fundamental constants ($k_B$ the Boltzmann constant, $h$ the Planck's constant), and is independent of the charge and the exchange statistics of the heat carrying particles. This quantum limit has been already experimentally realized for bosons [26, 27], fermions [28], and recently for a strongly interacting system – the fractional quantum Hall effect (with fractionally charged quasiparticles) [29]. In the latter study it was shown that in the presence of counter-propagating 1D modes, and in the limit of a long propagation length, the thermal conductance reflects the *net* number of the topological chiral modes (the number of *downstream* minus the number of *upstream*) – as predicted by the 'K-matrix' in the bulk [30].

Among the topological orders for the $v=5/2$ state (see above), the non-abelian candidates are predicted to conduct $n+1/2$ ($n=0-4$) units of the quantized heat, $J_Q$. The ½ originates from a neutral edge mode whose central charge is one half. This mode may be viewed as a Majorana edge mode of the superconductor of composite fermions. Moreover, each of the proposed topological orders has a different $n$ - implying a different quantized thermal conductance $g_Q$. Hence, measuring the heat conductance of the $v=5/2$ fractional state may provide an unambiguous determination of the nature of the state.

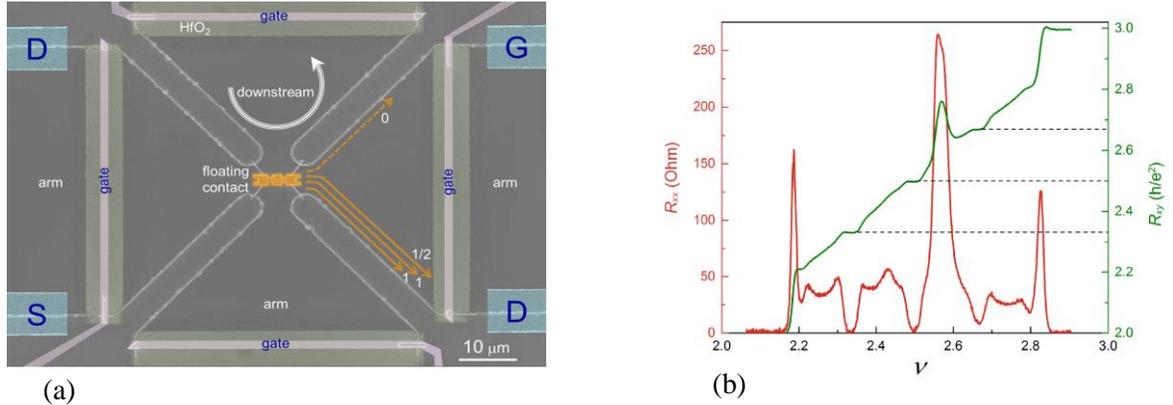

**Fig. 1. Device configuration and Hall data. (a).** The heart of the device For a description of the whole structure see Ref. 29. The small alloyed ohmic contact (with area some $12\mu m^2$) serves as a floating reservoir. Heated by the source (S) current, it injects outgoing currents into four arms. The effective propagation length of each arm is ~150μm. The thermal noise is measured in two arms (aided by two 'cold amplifiers'). Arms are pinched-off by negative charging of the surface gates (which are deposited on $HfO_2$). While the required voltage to pinch off a typical QPC was ~-10V, thus leading to severe hysteresis and instability, only ~-1V sufficed to pinch off the continuous gate, allowing a stable operation. We depict with the arrows the edge modes of the PH-Pfaffian order. The solid orange arrows are *downstream* charge modes, each carries $\kappa_0 T$ heat flux; the dashed orange arrow is an *upstream* Majorana-like mode, carrying $0.5\kappa_0 T$ heat flux (see Table in Extended Data). **(b)** The longitudinal and transverse resistances of the first-excited Landau level, as measured in a different Hall bar (200μm/L x 100μm/W),

Here we provide the crucial missing evidence in favor of a non-abelian order for the $v=5/2$ state; in particular we obtained $K=2.5\kappa_0$. This non-integer value not only confirms the existence of a neutral

Majorana fermions mode of central charge 1/2, but also agrees with the edge structure of the non-abelian PH-Pfaffian state [18, 19].

Our experimental setup, with its *heart* shown in Fig. 1a, is similar, in principle, to our previously studied configuration [29]; hence, we only elaborate briefly on it here. The structure is composed of a four-arm configuration (formed by chemical etching), with a DC input power provided by an impinging current in one of the arms at a small floating reservoir (an ohmic contact with area $\sim 12\mu m^2$). Each arm is crossed at its width with a continuous metallic surface gate (isolated from the sample's surface by 5nm thick $HfO_2$), some 30μm away from the floating reservoir. With $I_S$ being the source current, the impinging current on the floating reservoir is $I_{in}=t_1 I_S$, where $t_1=v_{gate}/v$ is the transmission coefficient of the source's arm gate ($v_{gate}$ the filling under the gate). The outgoing current (equilibrated at the ohmic contact) splits into the $N$ arms, with the dissipated power in the reservoir $\Delta P = P_{in} - P_{out} = 0.5 I_{in} V_S (1-N^{-1})$ – assuming an equal number of channels in each arm. The reservoir reaches thermal equilibrium at temperature $T_m$, when the dissipated power equals, ideally, to the outgoing power carried by the phonons (to the bulk) and by the edge modes; namely, $\Delta P = \Delta P_{ph} + \Delta P_e$. Consequently, the chiral 1D edge modes are expected to carry heat $\Delta P_e = 0.5 \cdot n_{tot} \cdot K \cdot (T_m^2 - T_0^2)$, where $n_{tot} K$ is the overall thermal conductance coefficient of $N$ arms, and $T_0$ is the electron temperature in the contacts. In turn, the heat flux carried by phonons is expected to obey $\Delta P_{ph} = \beta (T_m^5 - T_0^5)$ [31].

The temperature $T_m$ was determined by measuring the thermal noise carried by the current from the floating reservoir. We emphasize that due to the chirality of the edge modes and a conservation of the low-frequency current fluctuations, the noise in the electric current reflects the temperature of the floating contact. In practice, a cascade of a 'cold' (at 4.2K) and room temperature (RT) amplifiers amplifies the noise signal, which is measured by a spectrum analyzer. For example, the voltage gain of the 'cold amplifier', calibrated via thermal and shot-noise measurements, and normalized to the 30kHz bandwidth was ~9, with an input referred noise 300pV-Hz$^{-1/2}$; the gain of the RT amplifier was 200, with input referred noise 0.5nV-Hz$^{-1/2}$. The voltage fluctuations in the drain, $S_\upsilon$, passed an *LC* circuit at the mixing chamber ($f_0$~695kHz and BW=$\Delta f$=30kHz), and were converted to current fluctuations via $S_{th} = S_\upsilon G_H^2$, with $G_H = v e^2/h$.

Following the procedure described in Ref. 29, the temperature $T_m$ was plotted as a function of the dissipated power $\Delta P$. Additional contributions to the thermal conductance (*e.g.*, from phonons, bulk electrons), which depend only on $T_m$, were subtracted, $\delta P_{\Delta N} = \Delta P(N_i, T_m) - \Delta P(N_j, T_m)$, with $N_k$ being the total number of the open arms. This difference is directly related to the change of the heat flow due to a different number of the conducting arms. We plot a normalized coefficient for $N_i$-$N_j$ of open arms, $\frac{1}{\Delta N}\lambda_{\Delta N}=\frac{1}{\Delta N}\delta P_{\Delta N}/(\kappa_o/2)$ as a function of $T_m^2$, with the slope $K/\kappa_o$ for a single arm. We also show selected data for the total heat transport of $N_k$ open arms (without the above noted subtraction), which contains also heat flux that is not associated with the edge modes.

The grown structures in the experiments were different from the ones used in Ref. 29. The MBE grown GaAs-AlGaAs heterostructures that harbored the 2DEG were designed to screen effectively the ionized dopants. Consequently, 'relatively free' moving electrons that resided in the doping regions could contribute to the electrical and thermal conductance. We carried initial experiments with an extremely high mobility two-dimensional-electron-gas (2DEG), employing 'short period superlattice doping' (SPSL), with excess electrons in the doping layer [32]. The dark low-temperature mobility was $31\times10^6 cm^2$/V-s and the electron areal density was $3.1\times10^{11} cm^{-2}$. Clear Hall plateaus and $R_{xx}<10$ Ohms were observed for the range of fillings $v$=2-3 (see Extended Data). While the electrons in the doping layer are relatively localized (due to disorder), the layer still seemed to conduct heat. Testing the heat flow at $v$=2, where the thermal conductance is well understood, we found it to be higher than the correct value by some $3\kappa_o T$, and nearly independent of the number of conducting arms. Subtracting the heat flow as the number of arms changes, indeed led to the expected heat conductance in each arm (see Extended Data). Yet, the accuracy required in following experiments necessitated minimizing such an unwanted contribution.

Consequently, we developed a 'delta-doping' scheme in low Al mole-fraction of the AlGaAs layer (~23-25%). This led to shallower Si DX-like donor levels [33]. As in the SPSL scheme, excess doping was required to obtain full quantization of $v$=5/2 state. The shallower donor levels allowed a lower freezing temperature and thus a more efficient screening. However, complete localization was still achieved at mK temperatures. We conducted measurements in a sample with dark mobility $20\times10^6 cm^2$/V-s and the electron areal density $2.8\times10^{11} cm^{-2}$. While the span of the Hall plateaus and that of $R_{xx}$~0 was smaller (Fig. 1b), the sample seemed to have insignificant parallel thermal conductance.

The sample was recycled a few times to room temperature and back to low temperatures, and measurements were repeated at different temperatures and different magnetic fields (different fillings along the plateaus). With every set of measurements we tested: (*i*) Whether the source contact produces noise – and if present, the noise was subtracted; (*ii*) The equal division of the currents and the overlap of the Hall plateaus among the different arms; (*iii*) The reflection from the floating reservoir, which was found to depend on the recycling round and on $T_0$; (*iv*) The stability of $T_0$ during long measurements; and (*v*) The amplification of the amplifiers chain.

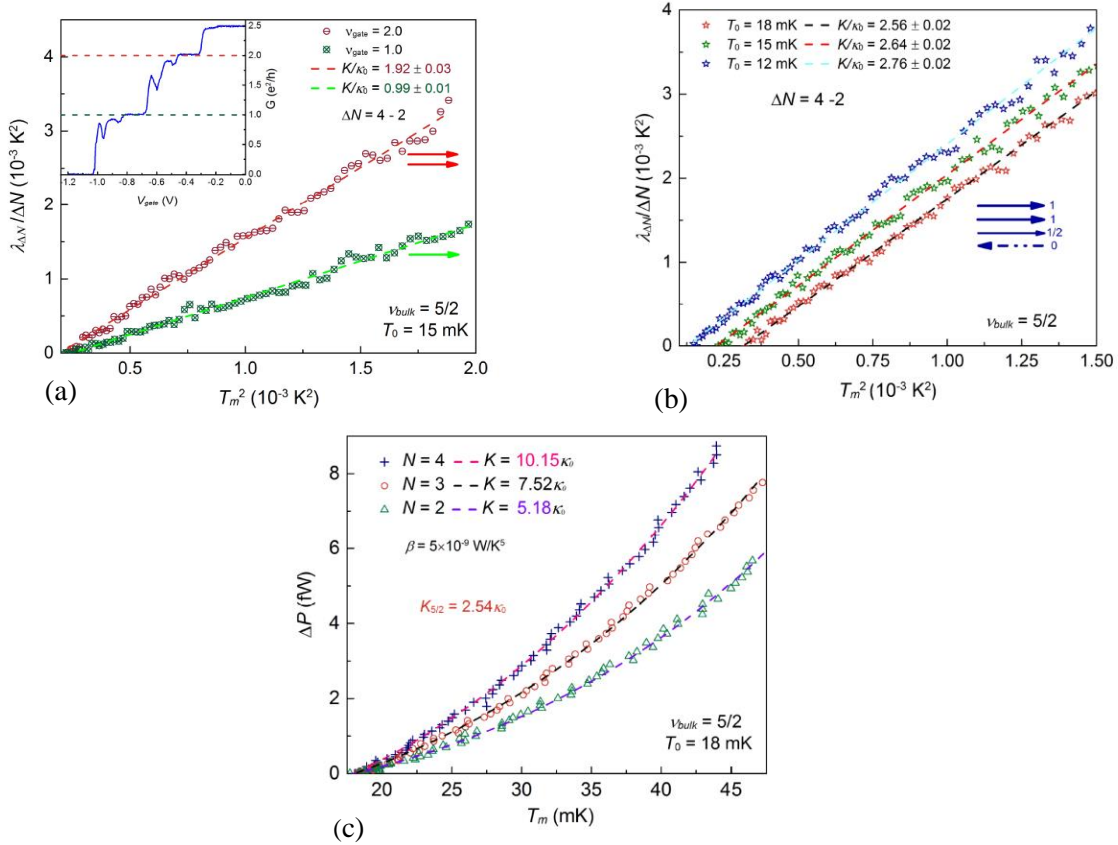

**Fig. 2. Heat flow of the two outermost edge channels in the $\nu=5/2$ bulk state.** (a) The normalized heat flow $\frac{1}{\Delta N}\lambda_{\Delta N}=\delta P_{\Delta N}/(\kappa_o/2)$, with $\delta P$ being the difference in heat dissipation for $N=4$ (four open arms) and $N=2$ (two arms are pinched off), as a function of $T_m^2$. Two cases are plotted: Two *downstream* edge modes per arm (two red arrows) and one *downstream* mode per arm (one green mode). The slopes are $K/\kappa_0\sim1.92$ for two edge modes and $K/\kappa_0\sim0.99$ for one edge mode. **Inset:** The transmission of a typical gate, allowing two or one edge modes in each arm. (b) Similar measurements in the $\nu=5/2$ state, with four and two arms fully transmitting. Representative measurements were performed at three different electron temperatures $T_0$. Here, the slopes indicate a weak temperature dependence of the thermal conductance, as it increases with lower temperatures. The arrows describe the idealized edge modes structure (for a long propagation length). Black arrows are *downstream* charge modes, each carries $\kappa_0 T$ heat flux; dashed pink arrow is upstream Majorana-like mode, carrying $0.5\kappa_0 T$ heat flux. (c) Total power dissipation with $N=4$, 3, and 2, is plotted as a function of $T_m$. A fit of $K$ in $\Delta P=0.5\cdot N\, K\cdot(T_m^2-T_0^2)+\beta\,(T_m^5-T_0^5)$ is determined, assuming the measured 'phonon coefficient', $\beta$. An average $K=2.54\kappa_0$ is found.

As detailed below, a few different methods of data analysis were employed (in different fractions), in order to minimize the uncertainty in our conclusions.

We start with the measurements performed in the $v=5/2$ state. Over the period of repeated measurements, the electrons hovered initially around 18-20mK (with fridge temperature ~10mK), and later around 12-14mK (with fridge temperature <6mK). In different cool-downs, the measured unintentional reflections from the floating small contact varied from a negligible amount to ~3%; yet, we did not find any significant difference in the acquired data in all cases. We measured first the heat conductance of the outmost integer edge modes in the $v=5/2$ state (known to be an integer multiple of $\kappa_o T$). This measurement verifies the integrity of our measurements (*e.g.*, the amplification and the electron temperature). This was done by tuning all the gates' voltages to either the first or the second conductance plateaus in Fig. 2 (namely, $v_{gate}=1$ or 2). We plot the normalized coefficient of the heat conductance, $K/\kappa_o$, of a single arm, as it is expressed in the slope of the normalized $\lambda_2/2$ ($\Delta N=4-2$) as a function of $T_m^2$ (Fig. 2b). We find an excellent agreement with the expected thermal conductance for a single edge mode ($v_{gate}=1$), as well as for two edge modes ($v_{gate}=2$).

In Fig. 2b we present the most important results of this work. With all arms gates unbiased, the heat was carried away from the floating reservoir by the total number of the topological edge modes of the $v=5/2$ state. With the average length of each arm being ~150μm, one expects nearly full equilibration of the counter-propagating modes (see also the discussion of $v=8/3$ state below). As before, we plot the normalized $\lambda_2/2$ ($\Delta N=4-2$) as a function of $T_m^2$ at three different electrons temperatures. We find a slight increase in the heat conductance as the temperature decreases, with $K_{5/2}=(2.56-2.76)\kappa_o$. Such temperature dependent increase is not surprising since the equilibration length increases at lower temperatures [29].

We also plot the total heat dissipation $\Delta P$ for different $N$ arms as function of $T_m$ (Fig. 2c). In this analysis we use a constant phonons prefactor (we find it for our small contact, $\beta$~5x10$^{-9}$W/K$^5$), with the plotted dotted lines are of $KT_m^2$ (that fit best the data points) for each $N$. The average $K_{5/2}=2.54\kappa_o$ per open arm.

It is interesting to look also at the two accompanying fractional states, the simpler $v=7/3$ state, and the hole-like $v=8/3$ state. In Fig. 3a, we repeat similar measurements to those performed at the $v=5/2$ state (in Figs. 2b & 2c). The 'subtraction' process leads to $K_{7/3}=2.86\kappa_o$, which is effectively $K_1=0.95\kappa_o$ per single edge (as the integer and 1/3 modes share the same heat conductance [29]). For the hole-like state we find $K_{8/3}=2.11\kappa_o$.

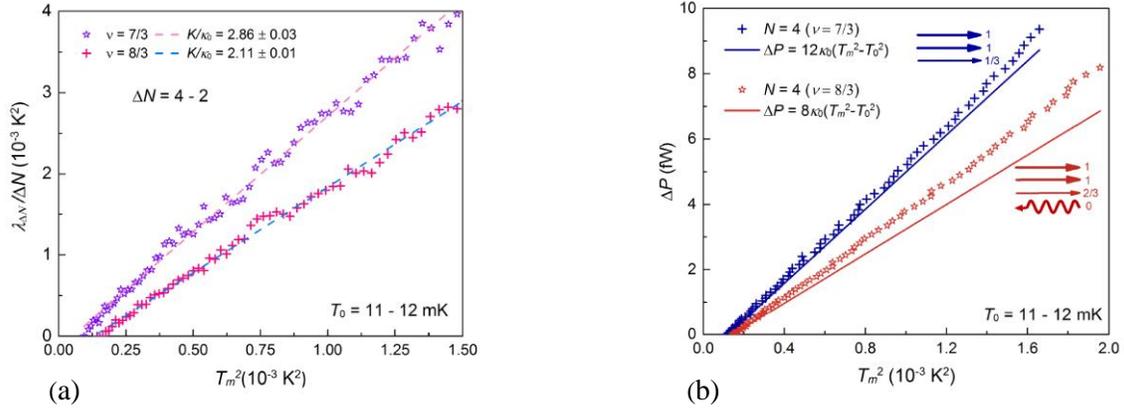

**Fig. 3. Normalized heat conductance at $v=7/3$ and $v=8/3$.** (a) A similar measurement to that described in Fig. 2a for the two fractional states in the first-excited Landau level near $v=5/2$. With the expected heat conductance coefficient in the $v=7/3$ state is $K=3\kappa_0$, we find a good agreement with $K=2.86\kappa_0$. In the $v=8/3$ state, with an *upstream* propagating neutral mode, one expects $K=2\kappa_0$ for an infinitely long propagation length. The observed thermal conductance coefficient is somewhat larger – as was also observed before in the fractional $v=2/3$ liquid [29]. (b) The total dissipated power is plotted as a function of $T_m^2$ (without data manipulation), for the two states in (a). Arrows describe the edge structure of each of the states. The actual power flow in the $v=7/3$ agrees well with the expected one at the lowest temperatures ($T_m<35$mK). At higher temperature the heat flow is higher (in both states) due to the added phonon contribution and lack of equilibration (in the $v=8/3$ state).

We present now also raw data, without any data analysis. The total heat dissipation for $N=4$ is plotted as function of $T_m^2$ (Fig. 3b). For the $v=7/3$ state, a deviation from $K_1=\kappa_o$ per mode (and $K_{7/3}=3\kappa_o$ per arm) is evident only when the temperature exceeds some 30mK, due to increased phonon contribution. Yet, for the $v=8/3$ state, the deviation from the expected thermal conductance of $2\kappa_o$ (for long distances) is quite apparent. Here, the phonon contribution and the non-equilibrated heat transport (at $T_m^{avaeage}\sim20$mK) both contribute to the larger deviation of the thermal conductance.

In Fig. 4 we summarize our results for the thermal conductance coefficient in the $v=5/2$ state. The summary is of some fifteen measurements (repeated ones are not shown), conducted over a two-month period, with multiple recycling to room temperature, at different electron temperatures, and at three fillings around $v=5/2$. The data points out to $K_{5/2}=2.5\kappa_o$.

Composite fermions and the K-matrix formalism provide a powerful framework for the understanding of the fractional quantum Hall effect in the lowest Landau level. Almost all the quantum Hall states in the lowest Landau level are believed to be integer quantum Hall liquids of composite fermions [34]. Our recent results on thermal transport in the lowest Landau level [29] strongly support that picture. The first excited Landau level poses a greater challenge. The nature of the $v=5/2$ and $v=12/5$ states has long been a puzzle. Competing proposals were also made for the $v=7/3$ and $v=8/3$ states [35, 36]. However, experiments show the quasiparticle charge of $e^*=e/3$ at both filling factors [37, 21]. A topologically protected upstream neutral mode is present at $v=8/3$ and absent at $v=7/3$ [36]. Those results and our current results for the thermal conductance of these two states are compatible with composite fermion liquids, similar to those at $v=2/3$ and $v=1/3$.

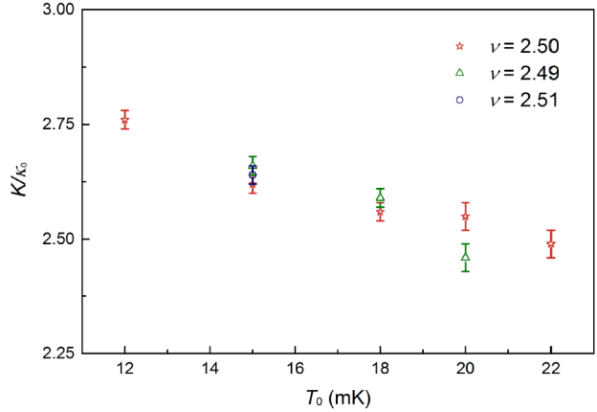

**Fig. 4. Summary of the normalized thermal conductance coefficient for the $v=5/2$ state.** We plot here the average $K/\kappa_0$ as a function of the temperature at three different fillings near $v=5/2$. A clear tendency of an increased thermal conductance at lower temperatures is visible. Such dependence is expected, since the equilibration length is expected to increases at lower temperatures [29]. Note that some fifteen measurements have been conducted, with most of them at $T_0$=15-22mK, with $K/\kappa_0$=1.55-1.65.

As mentioned in the introduction, theories of the $v=5/2$ are classified [38] to two types of topologically ordered states: abelian (such as 331, K=8, 113, and anti-331), with an integer quantized thermal Hall conductance and non-abelian (such as $SU(2)_2$, Pfaffian, PH-Pfaffian, Anti-Pfaffian, *etc.*) with half-integer thermal Hall conductance. [see Table in Extended Data].

We argue now that the observed heat conductances coefficient at $v=5/2$, being (2.54-2.56)$\kappa_0$ at 18mK, 2.64$\kappa_0$ at 15mK, and 2.76$\kappa_0$ at 12mK, fit well with the expectations of the non-abelian PH-Pfaffian state, having $K=2.5\kappa_0$. As the equilibration length is expected to increase at lower temperatures in states that support upstream modes, like $v=2/3$, $v=8/3$, and $v=5/2$ [29, 39], one would expect the apparent thermal conductance in the $v=5/2$ state to increase with lowering the temperature. Note that we did not find (in all our measurements) a thermal conductance coefficient that dipped below $K=2.5\kappa_0$. This resembles the behavior of the thermal conductance coefficient we find in the $v=2/3$ and the $v=8/3$ states; being always above the expected one for infinite edge length. The consistent results in the multiple measurements conducted in the known and the 'unknown' states, provide sufficient confidence in our conclusions.

While numerical works give support to two close relatives of the PH-Pfaffian state: the Pfaffian and anti-Pfaffian liquids [4, 11, 12], they generally do not focus on the inclusion of disorder. Addition of such effects seems to be essential for the comprehension of the observation we report here.

Our findings of half-integer thermal Hall conductance establish the non-abelian nature of the ν=5/2 liquid, making it the first state of matter experimentally shown to be non-abelian.

**Author contributions**

M.B fabricated the devices. M.B and M.H. designed the experiment, preformed the measurements, did the analysis and guided the experimental work. D.F., Y.O., & A.S. worked on the theoretical aspects. V.U. grew the actual 2DEG heterostructures. All contributed to the write up of the manuscript.

**Acknowledgments**

M.B. acknowledges the help and advice of Yaron Gross for suggestions regarding fabrication processes. M.B also acknowledges Y. C. Chung, H. K. Choi, R. Bhattacharyya and Ron Melcer for their help. M.H. acknowledges the continuous support of the SMC staff, which without them this wok would not be possible. M.H. acknowledges the support of European Research Council under the European Community's Seventh Framework Program (FP7/2007-2013)/ERC Grant agreement No. 339070, the partial support of the Minerva foundation, grant no. 711752, the Israeli Science Foundation, and together with V.U., the German Israeli Foundation (GIF), grant no. I-1241-303.10/2014. A.S and Y.O. acknowledge the European Research Council under the European Union's Seventh Framework Program (FP7/2007-2013) / ERC Project MUNATOP, the DFG (CRC/Transregio 183, EI 519/7-1), and the Israel Science Foundation. Y.O acknowledges the Binational Science Foundation (BSF). D. E. F's research was supported in part by the National Science Foundation under Grant No. DMR- 1607451.

# Extended Data

*MBE grown heterostructures*

The MBE growth techniques for heterostructures exhibiting the fragile $v=5/2$ fractional were developed during the last decades. Poor correlation was found between zero-field mobility and the $v=5/2$ state's energy gap. The key factor influencing the robustness of the state is found to be the spatial correlations among the charged scatters in the doped layers [32]. The highest quality (largest gap) was observed when using the SPSL doping scheme and by controlled illumination of samples at low temperatures. [40,41,42]. However, fabricated devices exhibited poor temporal stability as well as gate hysteresis effects making it virtually impossible to conduct conclusive experiments [42]. In this work we used two types of structures, characterized and operated in the dark, shown schematically below.

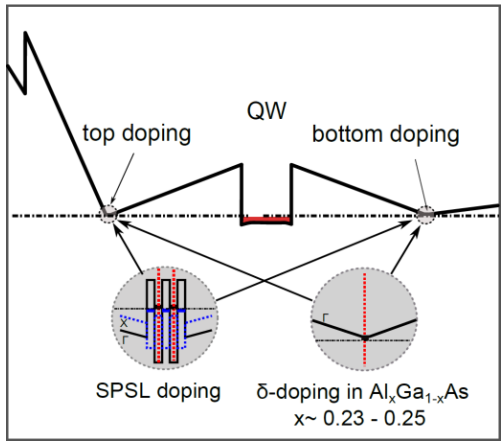

**Figure 1. Details of the growth structure.** Schematics of the conduction band in the MBE grown structures that are studied. The SPSL doping scheme comprises $\delta$-Si doping planes placed in narrow GaAs QWs; however, the thickness of GaAs and AlAs QWs in SPSL is chosen in such a way that the X-band minimum of the AlAs layers resides below the Γ-band minimum of the GaAs. This structure doesn't suffer from an added significant bulk heat conductance.

*Longitudinal resistance of high mobility SPSL MBE grown material*

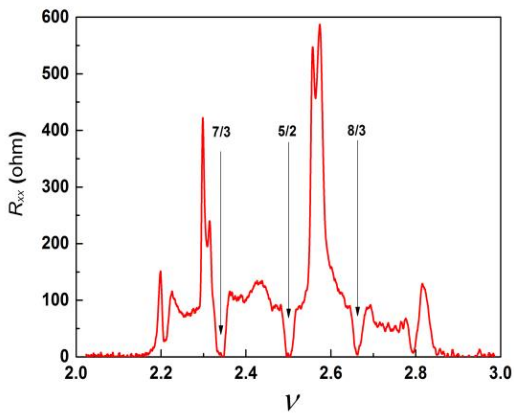

**Figure 2. Longitudinal resistance of SPSL material.** Longitudinal resistance measured in a Hall bar with effective lengths, 100μm wide, and 200μm long. Fractions here are more pronounced than in the $\delta$-Si doping structure. Yet, the structure suffers from added bulk conductance (see main text).

*Showing added thermal conductance in the SPSL structure*

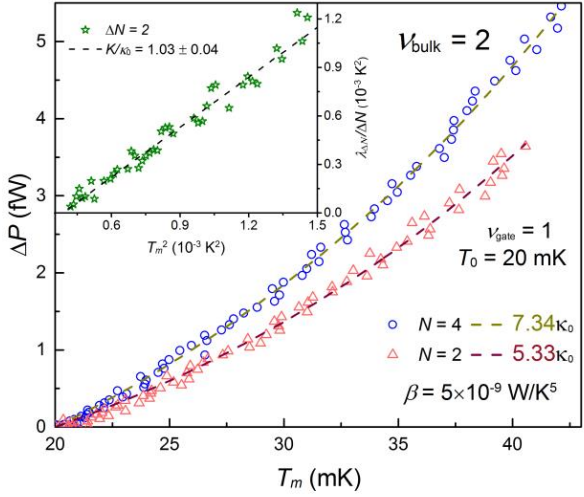

**Figure 3. Thermal noise analysis at *ν*=2 for SPSL structure.** Dissipated power in the floating reservoir is plotted as a function of $T_m$ for a different number $N$ of participation arms, with one edge channel allowed to flow in each arm (controlled by the surface gates). Dashed curves show the one-parameter fit of α, where $\Delta P(\alpha \kappa_0 T^2, \beta T^5)$, with β given (the average value deduced from all the experiments). Note the added parallel heat through the bulk ($7.34\kappa_0$ at $N$=4 instead of $4\kappa_0 T$, and $5.33\kappa_0$ at $N$=2 instead of $2\kappa_0$). **Inset**: Subtracting the contributions of different $N$ arms cancels the added phonons and bulk contributions. The fit line leads to an average thermal conductance per channel $g_Q=(1.03\pm0.04)\kappa_o T$, which agrees with the expectations.

*A different analysis of the thermal conductance of the ν=7/3 and ν=8/3 states*

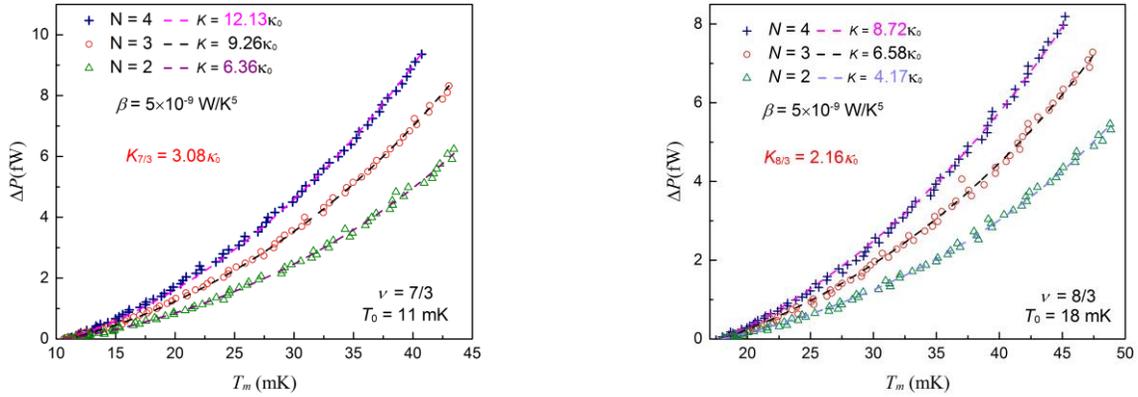

**Figure 4. Thermal noise analysis at *ν*=7/3 and *ν*=8/3.** Standard analysis (see main text), without subtracting the number of participating arms, with the average measured phonon contribution (β). A good agreement with the expectation is obvious. Note the added thermal heat in *ν*=8/3, which results due to the lack of full thermal equilibrium between the *downstream* and *upstream* modes (the length of the arms is not sufficiently long).

*Upstream neutral modes in v=5/2 and v=8/3*

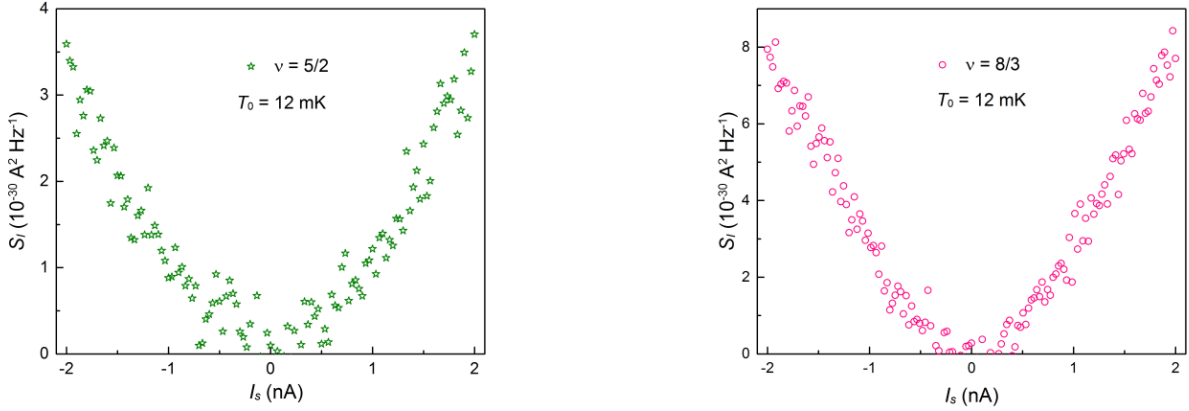

**Figure 5. Thermal noise due to upstream neutral edge channels, at *v*=8/3 and *v*=5/2.** The noise is measured at an upstream floating contact, which is connected to the 'cold' amplifier (with respect to ground). Such upstream noise is not measured in particle-like states.

*Possible orders predicted for ν=5/2 state*

**Figure 6. Table of nine possible orders predicted for $\nu=5/2$ state [16].** In this table we depict the edge structure of some of the leading candidates for the many body state of a fractional quantum-Hall state at $\nu = 5/2$ and their expected quantized thermal Hall conductance $K$ in units of $\kappa_0 = \frac{k_B^2 \pi^2}{3h} T$ with $k_B$ and $h$ Boltzmann's and Planck's constants respectively and $T$ the temperature. A right double line arrow denotes a downstream edge mode of a fermion with charge $e^* = e$ and $K = 1$. Right and left solid line arrows denote a *downstream* and *upstream* fractional charge mode with $e^* = e/4$ and $\kappa = 1$. The wavy line denotes a neutral mode with zero charge and $K = 1$, and finally the dashed line denotes a Majorana mode with zero charge and $K = 1/2$.

On the left (blue) part of the table we depict the abelian states with an integer $K$ and on the right (red) the non-abelian states with half integer $K$. We arrange the table from top to bottom with decreasing $K$. In all states the first Landau level is fully occupied so that to get the full edge structure we have to add two *downstream* fermions (depicted in two right black double line arrows in the left top part of the table). The $\nu = 5/2$ state can be constructed in a particle-like manner starting from $\nu = 2$ and adding fractional, neutral and/or Majorana modes, or in a hole-like fashion starting from $\nu = 3$ and adding modes moving in the opposite directions. This is emphasized in the edge structure of the hole-like anti-Pfaffian and anti-$SU(2)_2$ phases at the bottom part of the third column of the non-abelian states. *Downstream* and *upstream* charge modes as well as *downstream* and *upstream* neutral modes on the same edge equilibrate eventually. The final state of the edge, after equilibration, is depicted in the very right column of each row. The green line divides the particle-like and hole-like states. Notice that after equilibration, all states have a single downstream charge mode, as they should, but only the PH-Pfaffian is identical to the Anti-PH-Pfaffian – making it particle-hole symmetric.